\documentclass[journal]{vgtc}                % final (journal style)
\ifx\pdfoutput\undefined%              % if we use pure latex
  \usepackage{graphicx}                % allow us to embed graphics files
  \DeclareGraphicsExtensions{.eps}     % for pure latex we expect eps files
\else%                                 % else we use pdflatex
  \pdfoutput=1\relax                   % create PDFs from pdfLaTeX
  \usepackage{graphicx}                % allow us to embed graphics files
  \DeclareGraphicsExtensions{.pdf,.png,.jpg,.jpeg} % for pdflatex we expect .pdf, .png, or .jpg files
\fi%

\graphicspath{{figures/}{pictures/}{images/}{./}} % where to search for the images

\usepackage{microtype}                 % use micro-typography (slightly more compact, better to read)
\PassOptionsToPackage{warn}{textcomp}  % to address font issues with \textrightarrow
\usepackage{textcomp}                  % use better special symbols
\usepackage{mathptmx}                  % use matching math font
\usepackage{times}                     % we use Times as the main font
\usepackage{cite}
\usepackage[final,multiuser,marginclue,nomargin,inline,index]{fixme}
\usepackage[usenames,dvipsnames]{color}
\definecolor{grahammagenta}{RGB}{255,0,255}
\definecolor{ludoorange}{RGB}{220,165,25}
\definecolor{msblue}{RGB}{35,135,200}
\definecolor{ttwgreen}{RGB}{75,135,73}
\fxusetheme{colorsig}
\FXRegisterAuthor{tm}{atm}{\color{red}TM}
\FXRegisterAuthor{ttw}{attw}{\color{ttwgreen}TTW} % creates ttwnote, ttwwarning, ttwerror, ttwfatal commands
\FXRegisterAuthor{ms}{ams}{\color{msblue}MS}
\FXRegisterAuthor{mk}{amk}{\color{blue}MK}
\FXRegisterAuthor{gj}{agj}{\color{grahammagenta}GJ}
\FXRegisterAuthor{ludo}{aludo}{\color{ludoorange}LUDO}
\makeatletter
\renewcommand*\FXLayoutInline[3]{%
  {\@fxuseface{inline}\ignorespaces[#3 \fxnotename{#1}: #2]}}
\makeatother

\newcommand{\tone}{T1-Setup}
\newcommand{\ttwo}{T2-Validate}
\newcommand{\tthree}{T3-Identify}
\newcommand{\tfour}{T4-Default}
\newcommand{\tfive}{T5-Share}
\newcommand{\tsix}{T6-Improve}

\title{cellPACKexplorer: Interactive Model Building for Volumetric Data of Complex Cells}

\author{M. Schwarzl,  L. Autin, G. Johnson, T. Torsney-Weir and T. M{\"o}ller}
\authorfooter{
\item M. Schwarzl is with University of Vienna, Austria. E-mail: magda.schwarzl@gmail.com.
\item L. Autin is with The Scripps Research Institute. E-mail: autin@scripps.edu.
\item  G. Johnson is with University of California, San Francisco (UCSF). E-mail: 	Graham.Johnson@ucsf.edu.
 \item T. Torsney-Weir is with University of Vienna, Austria. E-mail: thomas.torsney-weir@univie.ac.at.
\item T. M{\"o}ller is with University of Vienna, Austria. E-mail: torsten.moeller@univie.ac.at. %MEMBER IEEE??
}

\abstract{Given an algorithm the quality of the output largely depends on a proper specification of the input parameters. A lot of work has been done to analyze tasks related to using a fixed model~\cite{sedlmair:2014} and finding a good set of inputs. In this paper we present a different scenario, \textit{model building}. In contrast to \textit{model usage} the underlying algorithm, i.e. the underlying model, changes and therefore the associated parameters also change. Developing a new algorithm requires a particular set of parameters that, on the one hand, give access to an expected range of outputs and, on the other hand, are still interpretable. As the model is developed and parameters are added, deleted, or changed different features of the outputs are of interest. Therefore it is important to find objective measures that quantify these features. In a \textit{model building} process these features are prone to change and need to be adaptable as the model changes. We discuss these problems in the application of cellPACK, a tool that generates virtual 3D cells. Our analysis is based on an output set generated by sampling the input parameter space. Hence we also present techniques and metrics to analyze an ensemble of probabilistic volumes.} % end of abstract


\CCScatlist{ % not used in journal version
 \CCScat{H.5.2}{User Interfaces}{User-centered design};
}

\vgtcinsertpkg

\begin{document}



\maketitle
%-------------------------------------------------------------------------
\section{Motivation}
\label{sec:motivation}

A typical modelling process consists of a model setup, an optimization process and a validation of that model. This is true whether dealing with agent-based modelling~\cite{miksch:hal}, statistical modelling (such as regression, classification, or clustering)~\cite{Bishop:2006} or computational modelling~\cite{UQreport:2012}. However, if the parameter space is vast, or if the optimization function is qualitative (e.g. through visual inspection) then this model building process can be quite tedious. This is one of the first papers that focuses on the visual support of model building in the biological domain. The main motivation is to observe a very specific model building process and to show that visual support can tremendously speed up and help in model building.

In this work we focus on cellPACK\cite{johnson:2015}, an open-source framework designed to generate and refine geometric structures of whole cells. One of the main challenges is to combine results of thousands of biological studies that provide small bits of knowledge to reveal how whole cells or components of cells function. Efforts to combine the results of these studies into comprehensive multi-scale cells, especially cells that take structure, compartmentalization and crowding into account, are almost non-existent.
Combining all the information and details spread over multiple sources, cellPACK requires multiple input parameters. This makes it very difficult for the developers to validate the functionality of each individual parameter as well as influence each other.

The ultimate goal of the cellPACK developers is to provide a tool that can assist researchers.  The current setup of cellPACK requires the user to specify a number of input parameters to build virtual cells. These parameters influence the complex interactions among the various molecular ``ingredients'' (e.g proteins) for a particular packing to produce a final molecular cell. Most of the parameters arose from different small-scale studies and often compete with each other. As these interactions between different parameters are very complex it is hard to predict the output related to a specific input setting. For more complex cases it is even impossible. The computation time for cellPACK outputs can be up to several days 
%using the current Python implementation 
which further increases the difficulty of heuristically finding a fitting input parameter set. Further, a proper validation of the model often happens visually by comparing the result to microscopy images or prior experience.  The aforementioned problems are a major bottleneck for the development of cellPACK and to the community participation required for consensus shaping on the scale of whole cells. New approaches are needed to make cellPACK more robust, easier to develop, and easier to test.

We have developed cellPACKexplore as a step to make developing cellPACK easier and assist the developers of cellPACK in the ongoing development.
%We describe the tasks of the cellPACK developers as \textit{model building} in contrast to \textit{model usage}. In \textit{model usage} a given model, or algorithm, is tuned to produce the output of interest. The underlying model is not changed in that scenario. In \textit{model building} the  underlying model is analyzed, adapted and improved.
cellPACKexplore provides an interface for accessing cellPACK,
setting up cellPACK experiments and for analyzing and sharing
cellPACK outputs and experiments. For this work, we focused on aiding the developers in
improving the core packing capabilities of their model to better help them
select the crucial features (input parameters), hide less important ones from future users, and to find proper defaults for some others so future cellPACK users (e.g. biologists, illustrators) are able to
quickly create cellPACK outputs themselves. In addition cellPACK developers constantly improve their model
to create cells of cells that are even more consistent with the current
domain knowledge and microscopy image data. This requires the implementation of additional input parameters.
Overall, cellPACKexplore supports the \textit{building} of cellPACK which further is used to create structural cells.

We implemented cellPACKexplore supporting a new workflow to simplify the development of cellPACK.
Our target audience are the developers of cellPACK.
Our contributions include a detailed user, data, and task analysis comparing the \textit{model building} tasks of the cellPACK developers to \textit{model usage} tasks which have been explored more thoroughly in the visualization community~\cite{sedlmair:2014}.
%We are following the conventions of Munzner~\cite{munzner:2015}. 
%They have a very different background. % and building a tool that all of them could use was not an easy problem. 
During the design of our tool we discovered how important it is to adapt visual representation to the user (i.e.\ a solution that works for one user might not work for another user).
%\msnote{I am not sure how much visualization research has been done in that direction, I would say this is very important. Additionally I wanted to bring in this idea early, as it validates our design decisions - reviewers critiqued that we don't have fancy new designs}\ttwnote{very little. there's some work by derek toker but it's on certain layouts for certain personality types and I imagine the effect sizes are tiny}). 
On the one hand, we were aiming to speed up their analysis workflow. On the other hand, we were aiming to improve the understanding of the complex behavior of the underlying model with the hope of further innovation. 
Some tasks that arise in the process of improving cellPACK are (i) choosing the right parameters (include or exclude features of the model that improve the expressiveness) (ii) setting proper (default) values for them, (iii) understanding the sensitivity of a parameter to be able to explain and predict its influence 
%and provide this information for future cellPACK users to help them choose the right input configuration to create the desired output 
and finally (iv) asserting a correct output after new parameters are added and existing ones are modified. Task (iv) is an ongoing process as the developers of cellPACK would like to be able to add new findings in biological research to their model.

%this is of interest to future users of cellPACK

Some of the technical challenges to overcome were the ability to deal with hierarchical data with tens of parameters as well as the exploration of an ensemble of probabilistic volumes (or ensembles of ensembles) which emerge from sampling cellPACK's input parameter space. We apply a filtering approach to this ensemble and suggest new metrics to validate the probabilistic volumetric dataset (see~\autoref{sec:abstraction:data} and~\autoref{sec:design:derived}). Using our tool the developers of cellPACK were able to speed up the setup to large experiments from 30~min to 1~min and were able to analyze ensembles of hundreds of cellPACK outputs which they could not do before. It also revealed unknown behavior of their tool to them and helped them to validate the influence of input parameters on the generated outputs. Another advantage the cellPACK developers pointed out is that cellPACKexplore makes collaboration and sharing of experiments easier. 

%-------------------------------------------------------------------------
\section{Related Work}
\label{sec:relatedWork}

The developers of cellPACK aim to efficiently build cells. In doing so they need to understand effects on the range of possible
outputs of cellPACK as they add and modify input parameters to the packing operations of
cellPACK. Hence, cellPACKexplore combines parameter space
analysis~\cite{sedlmair:2014} of the input parameters with ensemble analysis of
the 3D output in order to support \textit{model building} conducted by the
developers. Consequently, our work spans visualization research on \textit{model
building}, parameter space analysis, and ensemble analysis.

\subsection{Model building}

The \emph{usage} of simulation and modeling environments through the exploration of input-output relationships has been explored quite a bit in
the visualization community (see Sedlmair et al.~\cite{sedlmair:2014} for an
overview). However, visualization tools to support \emph{model building} are not as
common. \textit{Model building} is an iterative process where one identifies some
deficiency in the code that generates the simulation output, makes changes to address it, and then
\emph{validates} that the code change produced the desired
effect~\cite{UQreport:2012, miksch:hal, Bishop:2006}.
%This process is supported at a low-level (close to
%the code) by standard debugging tools or through a test-driven development
%process~\cite{Beck:2003}. Visualization has been used to help support this
%low-level process.  For example, Heapviz~\cite{Kelley:2012} presents an
%interactive view of the object graph stored in the heap. 
%These methods are focused on how specific parts of the code function
%at specific times which is good for detecting crashes and coding errors. We
%want to focus more on understanding how input parameters affect the outputs.
%\ttwnote{Maybe cut this low-level debugging discussion for space reasons}

As cellPACK is
under constant development, the available parameters are constantly being extended.
One of the core questions is whether these parameters properly capture the range
of realistic cells or whether they might be redundant with little influence to the
final output generated. Our approach to help the developers answer such questions is to let
them visually inspect the influence of new parameters on the range of possible
outputs. This could be considered as a hybrid approach between code-level
debugging and visual model building.

Most of the model usage tools assist the user with parameterizing a fixed model (i.e.\ a fixed algorithm). 
These methods are usually tied to a specific model.
For example, in the context of segmented regression~\cite{Muggeo:2008}
and treed regression~\cite{Alexander:1996}, Guo et al.~\cite{Guo:2011} focused
on the development and evaluation of linear models on subsets of the data. This
approach was extended by M{\"u}hlbacher and Piringer~\cite{Muhlbacher:2013} to
include non-linear trend discovery. Likewise, McGregor et
al.~\cite{McGregor:2015} present a system for Markov decision processes.
CVVisual~\cite{Bihlmaier:2015} provides code snippets that can be introduced
into image processing source code to provide debugging-type visualizations.
%When developing cellPACK, our eventual goal is to deploy it to domain scientists
%that are not programmers. Therefore, we designed cellPACKexplore to explore the
%range of possible cell model outputs and assist with finding what parameters
%produce the desired models.\ttwnote{grrrrrrrr, I don't like these last 2 sentences but I wanted to reiterate what makes cellpackexplore different} \tmnote{sorry, imho these two sentences are  abit misleading, so I removed them }

\subsection{Parameter space analysis}

A crucial aspect of developing effective models is understanding the expressiveness of a model. Hence, cellPACKexplore assists in grouping the outputs of the model for various parameter combinations based on similarity (of the output). An examination of the parameter sets that have created these groups helps to reason about the importance of specific parameters.
%is focused on splitting the range of outputs into various categories and then examining which parameters and ranges thereof have an effect in creating that category. 
Sedlmair et al.~\cite{sedlmair:2014} 
identify these tasks as \emph{partitioning} and \emph{sensitivity} tasks.

Partitioning is often done using a clustering approach.   For example, Design
Galleries~\cite{Marks:1997} uses a distance metric to present a set of visually
distinct possible renderings of a scene. Likewise, Fluid
explorer~\cite{Bruckner:2010} clusters a set of fluid simulation animations
into animation segments that are then inspected. 

However, while they are convenient, both approaches to a clustered presentation of the model outputs showed deficiencies. Hence, other researchers adopted a more manual adjustment. For example, Paramorama~\cite{Pretorius:2011} is focused on finding which parameter settings
produce good segmentations based on manual inspection of the resulting images.
They group outputs in a hierarchical fashion based on input parameter
settings. Paraglide~\cite{Bergner:2013} enables a manual partitioning of the output space in order to draw conclusions on the input parameter values.
%addresses the issue of how many samples
%are required to generate all possible behaviors of the algorithm.  
An a priori partitioning scheme is not clear in the case of cellPACKexplore.
%, the issue is that the developers don't have a partitioning
%scheme a priori.  
In addition, the partitioning scheme might be refined as the algorithmic description (and therefore the underlying model)
changes. An evolving model description often brings new parameters with unknown effects.
Therefore, we support a manual partitioning of the data by letting the user filter on input
parameters and metrics computed on the output.

With cellPACKexplore, we rely
on the user to discover patterns within and between the histograms of input parameters
and metrics computed on the output. This is similar in approach to the star glyph overview
display of Guo et al.~\cite{Guo:2011}. In our case, the histograms were a visual
encoding that the cellPACK developers were familiar with.

%Visualization tools have so
%far focused on local sensitivity. For example, Guo et al.~\cite{Guo:2011} use 
%star glyphs and the user can discover sensitivity patterns.
%With cellPACKexplore, we take a more global sensitivity approach. Using 
%histograms allows us to show the sensitivity without the cellPACK developers
%having to learn a new visual encoding.

%\begin{itemize}
%\item Tuner~\cite{Torsney-Weir:2011}
%\item SA star glyph?~\cite{???}
%\item design galleries?~\cite{???}
%\end{itemize}

\subsection{Ensemble analysis}

%Test-driven development requires some sort of metric for a correct solution
%but in order to come up with these metrics. CellPACKexplore helps with this
%by showing the range of possible outputs and allows the user to classify
%groups of outputs by performance in order to help support this task.

One of the complexities of understanding the results of cellPACK is that the
packing algorithm is stochastic. 
In other words, for a particular parameter
configuration cellPACK produces an ensemble of volumes.
In cellPACKexplore we want to analyze these ensembles at two different levels.
We want to group a number of outputs into distinct sets of outputs based on
large-scale differences (see the previous section) and analyze the smaller-scale variations within these
sets. While there are approaches to ensemble analysis and approaches that treat ensembles as distributions (see Kehrer and Hauser for an overview~\cite{Kehrer:2013}), to the best of our knowledge there are no visualization approaches that deal with a hierarchy of ensembles as we do here.
%To our knowledge, while there have been visualization systems focusing on
%grouping multiple deterministic outputs and systems focused on analyzing a
%single stochastic output, no system focuses on both as we do in this work.

%%When analyzing multi-run data with deterministic outputs one usually wants
%%to group these into bins or categories.  This grouping is often based on
%%large-scale variation and is often handled with clustering
%%algorithms~\cite{Aggarwal:2013}. For example, Design
%%Galleries~\cite{Marks:1997} uses a distance metric to present a set of visually
%%distinct possible renderings of a scene. Likewise, Fluid
%%explorer~\cite{Bruckner:2010} clusters a set of fluid simulation animations
%%into animation segments that are then inspected.  The goal is to find the best
%%fluid simulation. In cellPACKExplore, we also tried clustering the outputs automatically but
%%discovered that finding a good metric to cluster on 
%%%and an adequate algorithm that works on that features 
%%was difficult as the developers wanted fine-grained control over creating
%%output sets. Hence we changed to a manual partitioning approach through
%%interactive filtering of input and output parameters (see~\autoref{sec:analyzeScreen}).
%%%In addition, none of
%%%these systems support stochastic output.\msnote{referring to systems of the
%%%previous sentence or input+output?}

For stochastic simulations, one usually examines the distribution of output
possibilities resulting from a single parameter configuration.  
%The most
%well-studied and perhaps intuitive method of thinking about distributions is in
%1D (i.e.\ scalars).  Correll and Gleicher~\cite{Correll:2014} survey methods for visualizing these
%1D distributions\msnote{@Tom: this does not refer to the distribution in the interface right?}. 
To help show these distributions, 
the notion of a boxplot was extended for curves
with the introduction of contour boxplots and curve
boxplots~\cite{Whitaker:2013, Mirzargar:2014}. These were used, for example,
for visualizing the range of possibilities of storm tracks. These work well
for showing distributions of 1D functions, but cellPACK produces two- or
three-dimensional outputs.
VAICo~\cite{Schmidt:2013}
considers a set of 2D images and computes the regions of difference of the
set, clusters them, and then gives the user controls to browse these
differences. %One can identify outlier areas in regions of the image.  
While the aim of VAICo is to identify pixel level differences in images we are looking at understanding structural volumetric differences in the set of (probabilistic) volumes.
%and is not ideal for
%finding large-scale differences in the set. 
%We look for these large-scale
%differences in cellPACKexplore to help reorganize the resulting 3D images into
%clusters of common structure.

cellPACK is primarily intended for 3D output.  Three-dimensional
objects are often represented as either voxels or parametric objects. For
voxel-based data, there are visualization methods such as probabilistic
marching cubes~\cite{Pothkow:2011} or MObjects~\cite{Reh:2013}.  One can also
animate between all 3D objects in an ensemble as in Ehlschlaeger et
al.~\cite{Ehlschlaeger:1997} or Lundstr{\"o}m et al.~\cite{Lundstrom:2007}.
While animation techniques will work on general 3D output, animation can
contribute to a higher cognitive load for users especially if the time axis in
the animation does not correspond to time in the data~\cite{Tversky:2002}. In
cellPACKexplore, we combined the intuitive notion of a 1D distribution with the
detail of 3D. We show 1D distributions of derived metrics with a
user-selectable view of a single projected 3D output.
% \msnote{@Tom: this can also stay right?} \ttwnote{why wouldn't it?}

%-------------------------------------------------------------------------
\section{cellPACK}
\label{sec:cellpack}

Since cellPACKexplore is designed to work with cellPACK we provide an introduction to cellPACK itself here.
cellPACK is an open-source biological software framework designed to assemble large-scale cells and cellular substructures from small-scale molecular building-blocks.
% cellPACK was initially developed to create figures for publications and communication in research and teaching. 
%The developers adapted their algorithm \ttwnote{a word is missing here, also, what was cellpack before? why didn't the parameters have biological meaning?} changed input parameters to give them a biological meaning. 
cellPACK was designed to combine data from all branches of biology into comprehensive cells. As parameters reflect real biological properties of proteins and their interaction cellPACK can be used for hypothesis generation and experimentation (imitating localizations and interactions), validation, communication, education and to view the mesoscale ($10^{-7}\ldots10^{-8}m$) with atomic-resolution detail. The ultimate goal is that cellPACK will be accessible to audiences without a technical background and serve as a structural and informatics foundation for broader projects~\footnote{\url{http://www.scripps.edu/newsandviews/e\%5F20150921/vmcc.html}}
\url{www.scripps.edu/newsandviews/e_20150921/vmcc.html} \msnote{@TOM: link doesn't work on click - but if you type the link it works - do you know how to solve this?} \ttwnote{I replaced it with the url encoded value thing and that works. Is that ok? It's weird. underscores should be ok according to stack overflow}
 and the new Allen Institute for Cell Biology, which aim to generate dynamic virtual representations of whole cells for predictive experimentation.

\begin{figure}[tb]
  \centering
  \includegraphics[width=.5\linewidth]{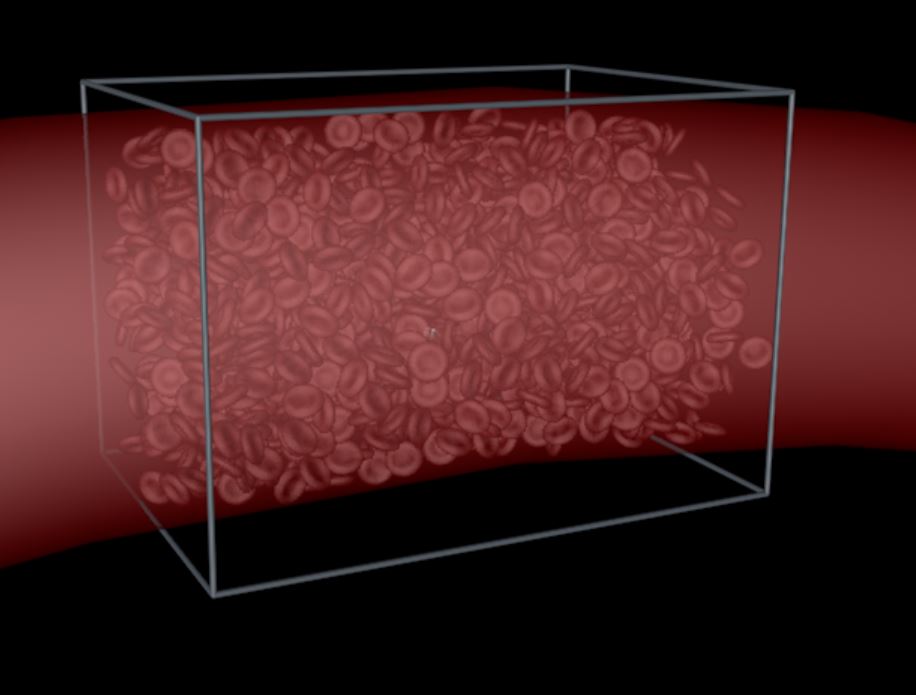}
  \caption{\label{fig:Autopack} A simulation of the red blood cell distribution (1413 cells) in a blood capillaries of radius 30$\mu$m and length 100$\mu$m built with cellPACK.}
\end{figure}

A cellPACK input file (called a recipe) contains a list of molecular building-block components (called ingredients) with behaviors (input parameters) that mimic biological constraints (e.g.\ attraction and repulsion). Given a molecular recipe, the construction of a quantitative 3D mesoscale cell requires solving a loose-packing problem. In biological systems, this includes packing soluble, membranous, and fibrous components with proper localizations and biologically relevant interactions. autoPACK serves as the core packing engine for cellPACK and provides a generalized solution to the loose-packing problem.
It is useful by itself for researchers from other domains as well (e.g. generating densely packed volumes or surfaces for engineering and computer graphics). It can be used, for example, to fill an architectural engineering shape with concrete aggregate in preparation for earthquake simulations, or it can fill an artery with blood cells at appropriate densities to generate a histological representation for a medical illustration \autoref{fig:Autopack}.
%-------------------------------
% HIV packing figure
%-------------------------------
3D shapes in autoPACK (and hence in cellPACK) have relatively simple agent behavior parameters that guide how they pack together (i.e. which position an ingredient takes in the volume). The cellPACK developers predict the output cell if a single ingredient is packed. If more than one ingredient type (i.e. different proteins or molecules) is specified each type has its own characteristic. The interaction (and sometimes competition) of input parameters (i.e. ingredients) coupled with stochastic variations can build up to generate 3D cells with emergent complexity that humans cannot predict.

%-------------------------------
% HIV packing figure
\begin{figure}[tb]
  \centering
  \includegraphics[width=.99\linewidth]{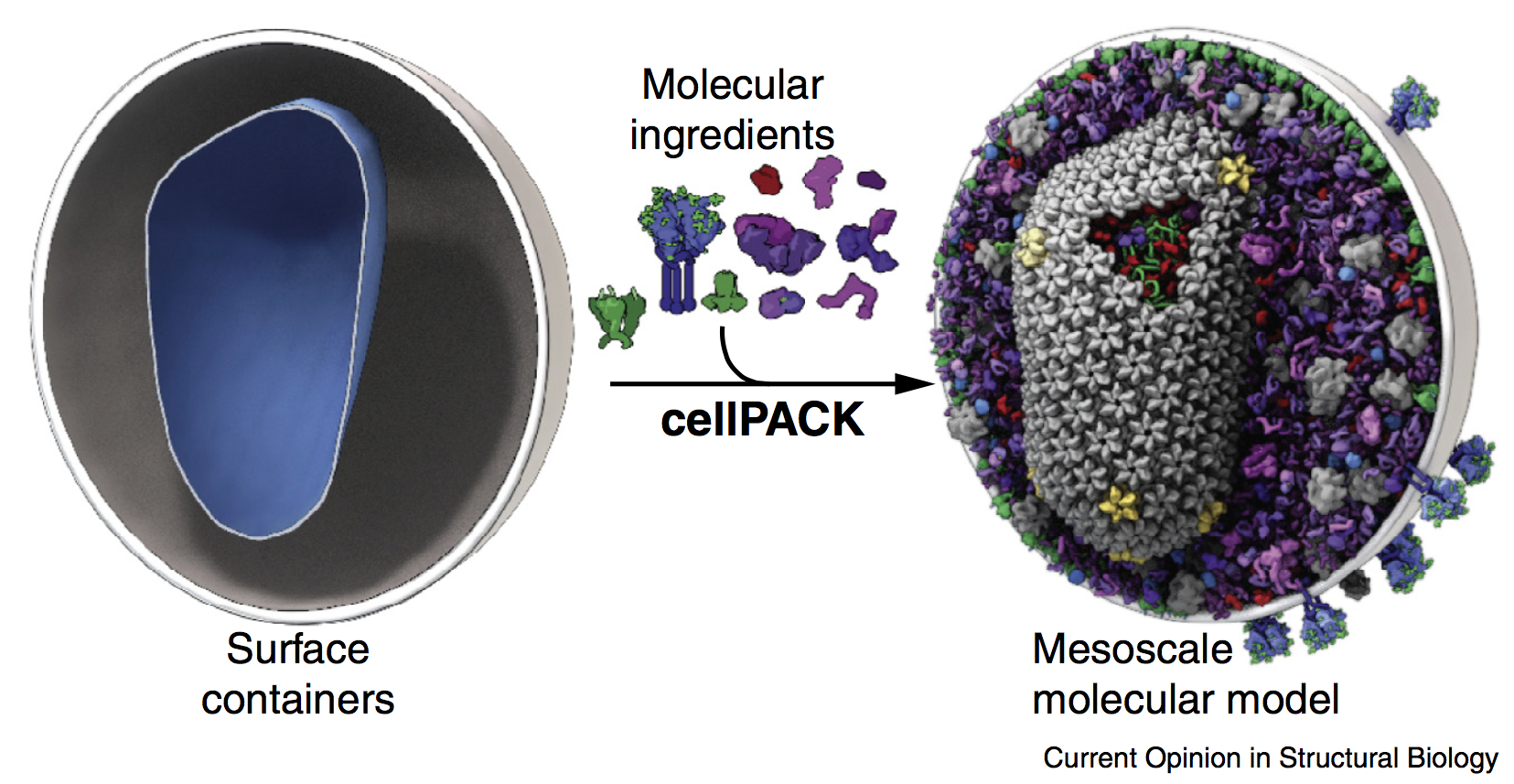}
  \caption{\label{fig:HIV}Example of a cellPACK output that results from packing an HIV recipe. Left: the empty packing volume. Middle: different ingredients (proteins) to be packed. Right: one (of many) stochastic packing results. Note the emergent complexity of the protruding green/blue ingredients, which packed with a bias towards one side of the spherical surface, as the result of several simple molecular building blocks interacting in a variety of localized manners. Figure used with permission \cite{johnson:2015}.}
\end{figure}
%-------------------------------

One major research focus of the developers of cellPACK is to create rigorous structural cells of the Human Immunodeficiency Virus (HIV) and the virus's interactions with human cells. These cells and the recipes used to build them help to unify and frame diverse knowledge into the context and scale of the entire virion. \autoref{fig:HIV} demonstrates an example of the packing problem for HIV. Simulations like this can impel hypothesis-driven research by integrating diverse data types, identifying gaps in knowledge, and exploring the ranges of cells that are consistent with current knowledge. Of course this requires a valid (according to current domain knowledge) underlying model. Developing such a model is one of the major goals of the cellPACK developers and requires them to validate, adapt and better understand cellPACK.
A second major focus of the cellPACK developers is on their core packing algorithms, namely autoPACK. The development of cellPACKexplore has been focused primarily around improving these core algorithms and explore the behavior and interactions of new input parameters. An approach of the cellPACK developers is to pack spheres with different radii into a box or on a plane to explore the variations in outputs using different input configurations (e.g. different parameter values for attraction between two sphere types).

cellPACK's input parameters can be split into two groups, general parameters and ingredient parameters.
\textbf{General parameters} influence properties of the packing algorithm affecting the whole cell. Two examples include the resolution of cellPACK's spatial tracking grid, and a variety of options for how the next point on the grid to be packed (assigned to an ingredient) is selected. These are equivalent to \textit{model parameters}~\cite{sant:2003} and do not provide biological information about a specific protein's behavior.

\textbf{Ingredient parameters} mimic the behavior of different real-world biological  ingredients, such as molecular proteins and can be set for each ingredient type independently. These parameters include a parameter that lists known binding partners for an ingredient and methods for how it should interact with each potential partner (i.e. attraction). Another example are  parameters that regulate the orientation of an ingredient (not meaningful if spheres are packed). These have been referred to as \textit{control parameters} in the literature~\cite{sant:2003}, as they are meaningful in the biological domain and contribute locally (position for each individual ingredient) to solve the general packing problem globally (build the whole cell). Ingredient parameters are of interest for hypothesis driven research. One can specify them and compare the generated output cell to images created with a microscope. If there are multiple input configurations (e.g. binding probabilities) that create a valid output more wet-lab experiments are necessary to gather missing information. New insights can then be incorporated into cellPACK.

%-------------------------------------------------------------------------
\section{Problem Characterization and Abstraction}
\label{sec:abstraction}

Based on weekly meetings with the developers of cellPACK over the course of ten
months, we analyzed their workflows and tasks and
%users of cellPACKexplore (i.e. the developers of cellPACK), 
characterized their data source (i.e. the
cellPACK model).
We then synthesized what we consider the main 
%, and the 
model building tasks conducted by the developers of cellPACK in
order to help them improve the understanding of their model and ultimately improve
cellPACK. In what follows we summarize the result of our analysis.

%%%%%%%%
%%% USER
%-------------------------------------------------------------------------
\subsection{cellPACK Developers}
\label{sec:abstraction:user}

The ultimate goal of the cellPACK developers is to make their tool accessible 
for a broad audience (e.g. biologists, illustrators). To achieve this goal they want to
improve the accessibility of cellPACK. Ideally, the complexity of the model which is tightly correlated to the number of parameters, should be decreased. The remaining few parameters should then be well documented and get reasonable default values. Hence, it is crucial to
understand the influence of parameters on the model and the output it
produces as well as the possible interaction between different parameters. On top of that the developers have to make sure the parameters are following biological constrains. 

The collaboration with the developers of cellPACK started with the intent to
make cellPACK accessible to a large set of (non-technical) biologists. However,
we quickly realized that cellPACK is still under development and cell modelling by itself
is very complex. We therefore decided to focus on the two core developers of cellPACK as users of our tool, cellPACKexplore. 
Their goal is to compute realistic cells of
biological structures (e.g. HIV, Blood-Plasma) from smaller components (called
ingredients). Their approach is to extract information from various small-scale studies and develop
parameters and code that follow these restrictions. There models (e.g. HIV or Blood-Plasma cells) are refined iteratively
by changing the parameter settings (model-usage) or adapting a parameter's behavior and adding different parameters (model-building) until the
output produced by cellPACK corresponds to the observed image in the
microscope. By building a number of different specific cells, the developers
gain an understanding of the importance of specific parameters.

Designing a tool for the developers it was important to consider their in-depth knowledge about cellPACK. In comparison to other visualization tools, the developers do not just use their tool and fine-tune its parameters to create the output of interest but rather adapt the underlying model directly to work as expected and thereby adding and removing parameters from consideration. Hence, we found this to be one of the distinguishing aspects of \textit{model building} as opposed to \textit{model usage}.

Specifically, there are two developers that we engaged in, both have very
different approaches in validating the code which added some complexity in designing an interface that both of them could use. They come from very 
different backgrounds. One of them has a background in scientific illustration
and focuses more on the visual correctness. The other developer has a deeper
technical background and focuses more on derived statistical metrics to explore
the dataset. We aimed at building a system that supports both and discovered the importance of adapting the interface for targeted users.
In the end, We show both, visual summaries of outputs (\autoref{fig:analyzeWhole}c,
\autoref{sec:design:packing}) and statistical measures
(\autoref{fig:analyzeWhole}b, \autoref{sec:design:derived}). However, our interface adapts according to the needs of the user. We have found, that specifically for supporting the building of models, it is unclear a-priori which measures will be needed to summarize the outputs. In contrast, model usage environments typically have a good understanding which features of the data are of interest.
%This is also a specific requirement for model building. In contrast to model usage where the underlying data does not change users know ahead what features of the data they are interested in. 
In model building the underlying data changes and the features a user wants to inspect change with the data. Therefore the interface has to provide the capability to adapt to these changes.

%%%%%%%%
%%% DATA
%-------------------------------------------------------------------------
\subsection{Data}
\label{sec:abstraction:data}
Since we are focusing on model building rather than model usage, the real data analyzed in cellPACKexplore is the cellPACK model itself. To better understand, analyze and validate cellPACK, we support analyzing an ensemble of cellPACK outputs created by sampling the input parameter space.

cellPACKexplore takes multiple input configurations for cellPACK to compute multiple cellPACK outputs. For each input configuration, cellPACK's output consists of multiple 3D positions for each individual ingredient type. 
%Using the convention of Munzner~\cite{munzner:2015} this data is best described  as a \textit{spatial field}.
%We create an ensemble of these spatial fields by varying the input parameter settings for cellPACK. 
Each single parameterization of the cellPACK model produces a number of outputs by stochastically varying the naturally occurring variations in the biological cells (i.e. initializing the algorithm with different random seeds). This gives a two-level hierarchy to the ensemble: each parameter configuration generates a number of outputs which differ by the used initialization for the random number generator. We define a \textit{run} as the creation of $R$ different simulation outputs by re-running the simulation $R$ times with the same parameter configuration but with a different random seed each time. A run results in a probabilistic volume for each input setting. In the interface all filters work on the level of runs as atomic units.
%, as it is related to one parameter set. 
We define an \textit{experiment} to be a subset of parameters that are varied over a range of parameter values.
% and its related outputs. 
An experiment consists of $N$ different runs by selecting $N$ parameter configurations at random over the set of chosen parameters. This results in $N$ \textit{sets} of $R$ \textit{results} giving a total number of $N \times R$ volumes.

The cellPACK input parameters have different data types. Some of them are categorical, e.g. specifying which algorithm should be used to handle intersections of ingredients. Others are numerical, e.g. influencing the binding probabilities between ingredient types. %Working on the code for a long time, the developers of cellPACK have changed many parts and often add new parameters and functions. New parameters can supersede or otherwise affect other parameters and the downstream results produced.

In addition to raw outputs, we introduce statistical metrics to describe the characteristics of the output ensemble. Although we anticipated that these metrics simplify the navigation through the huge number of output ensembles, the meaning of some of the metrics was difficult to make clear to the cellPACK developer with a visual background. He preferred to rely on the graphical rendering of the cell. To address this we show both the 3D outputs and the derived metrics.

%%%%%%%%
%%% TASK
%-------------------------------------------------------------------------

\subsection{Tasks}
\label{sec:abstraction:task}

%------------------------------------------
% Workflow figure
\begin{figure}[tb]
  \centering
  \includegraphics[height=.63\linewidth]{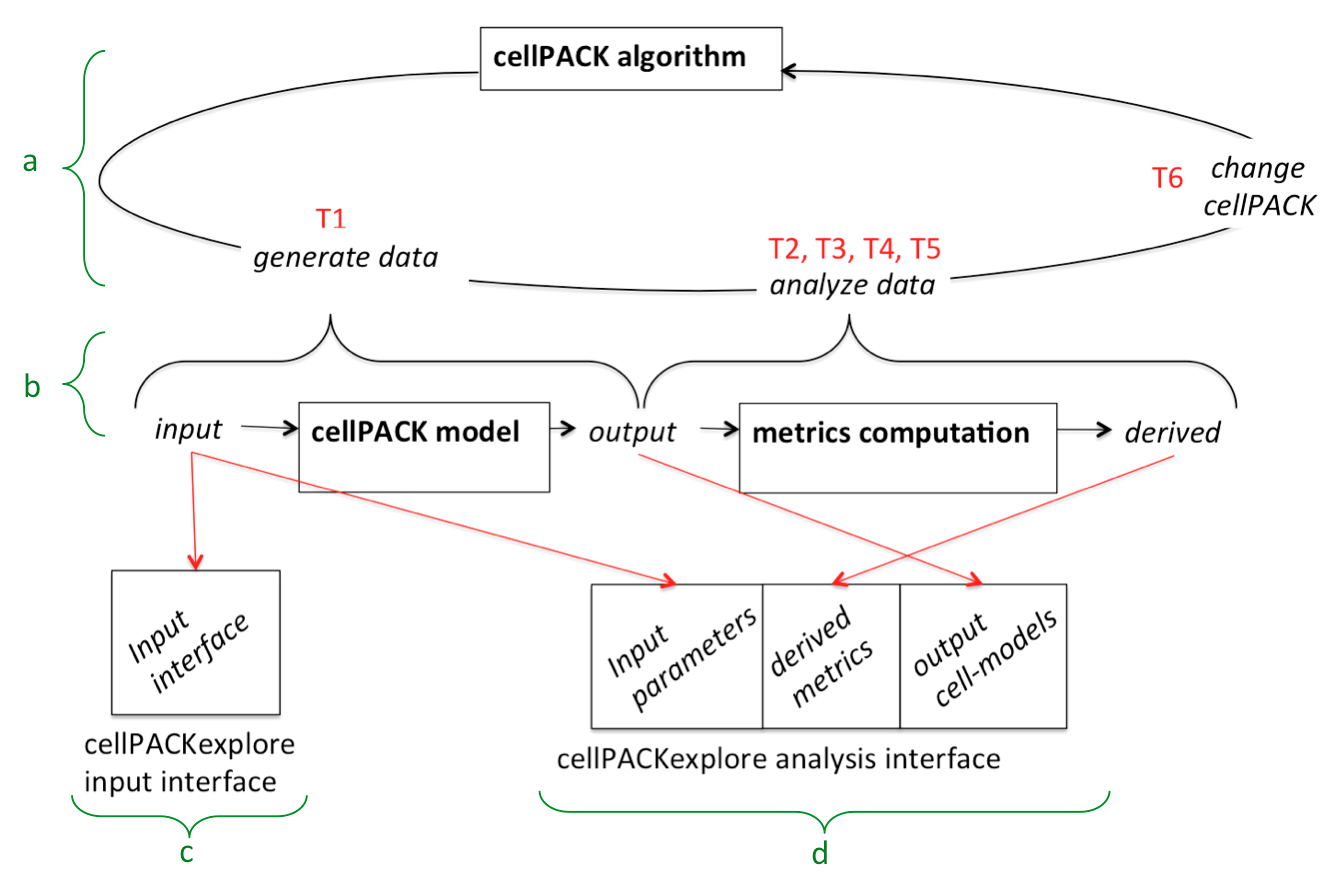}

  \caption{\
    The conceptual workflow of the developers of cellPACK using
    cellPACKexplore.  (a) The developer's mental representation of the
    workflow. They start with a version of cellPACK code, create
    a number of outputs to test it (\tone), and, based on their data
    analysis (\ttwo, \tthree, \tfour), discuss their results (\tfive) and create an improved version of the
    cellPACK code (\tsix). Prior to cellPACKexplore, generating and analyzing the
    data was done as a tedious manual process. (b) cellPACKexplore's pipeline
    guides the developers through the analysis process. (c) The setup of an experiment (see~\autoref{sec:inputScreen}) as well as (d) the analysis of the experiment (see~\autoref{sec:analyzeScreen}) are supported through visual interactive interfaces and improves their workflow.}
\label{fig:workflow} \end{figure}
%------------------------------------------

We start this section by
describing the current workflow of the cellPACK developers. Based on this we abstracted 
the tasks (\tone{} through \tsix) and propose an improved workflow using cellPACKexplore.
A conceptualization of our resulting new pipeline can be found in \autoref{fig:workflow}.

Currently, the developers use a simple \textit{trial and
error}~\cite{sedlmair:2014} strategy, exploring one run at a time by comparing it 
to current domain knowledge. They create a hypothesis of what the cellPACK outputs 
for specific parameter settings will look like and then run the model to verify 
their prediction. This approach works well for experiments with a single ingredient 
type. As soon as more than two ingredient types are packed, parameters start 
competing against each other which makes it impossible for them to predict the 
output. Before cellPACKexplore the cellPACK developers did not have an easy solution for 
this issue. After roughing out parameters and ranges of interest running one packing at a time 
they write custom scripts (\tone) to sample a subset of parameter ranges
(typically varying seeds and only 2 to 3 parameters). The outputs are then
analyzed visually (\ttwo) using barcharts in MS Excel and density maps that
require a suite of different 2D or 3D viewers. If this analysis leads to an insight
the underlying model (i.e. cellPACK) is changed (\tsix) and a new iteration starts.
A bottleneck in their current workflow is the computation time for a single cell. It can range
from a 100th of a second to several days for more complex recipes packing
millions of base pairs or proteins.
Another limitation is the communication overhead. One developer has a background 
in scientific illustration and usually asks the other developer to write scripts to generate measures he
is interested in. This requires communication between the developers and a possibility to
share data (\tfive) currently they are using a git repository.
One thing missing in their approach is the ability to analyze the complex interplay between
different parameters. Currently the developers do not have a tool to visualize these relationships. In this paper we focused first on visualizing parameters independently; new features to directly show interactions between parameters remain for future work.

Based on the cellPACK developer's current workflow and on their goals we 
abstracted the following tasks:

\textbf{\tone: Experiment setup:}
  This requires selecting a subset of input parameters, a range for
  sampling, and a decision on the number of samples to be generated (compare
  $N,R$ in \autoref{sec:abstraction:data}). Usually the technically-trained
  developer was responsible of creating runs and outputs as well as
  statistical summaries, while the other developer engaged in the validation
  with respect to biological (or other) ground truth.

\textbf{\ttwo: Model validation through output comparison:}
  When parts of the code are changed, it is important to make sure that the
    cellPACK model is still valid and produces correct results by comparing
    with microscopy data and current domain knowledge. This requires the
    analysis of the probabilistic volume ensemble set related to an
    \textit{experiment}. cellPACK outputs are checked to ensure they satisfy
    several statistical constraints like concentration of ingredients and
    distribution over the volume.

\textbf{\tthree: Identify parameters to be exposed:}
  Parameters that greatly impact the range of 3D outputs should be exposed to the future users of cellPACK. However, too many parameters could overwhelm a new user with unnecessary complexity and hurt the adoption of cellPACK. In addition, packing parameters might not be intuitive to non-technical users. Therefore, the developers of cellPACK have to make a careful selection and need to understand the behavior of different parameters and their interactions.

\textbf{\tfour: Identify reasonable default values for other (hidden) parameters:}
  After identifying which parameters to expose, the developers of cellPACK
  need to decide what are reasonable defaults for the remaining parameters.
  These default values should produce
  accurate results without additional configuration.

\textbf{\tfive: Share results:}
  As the developers of cellPACK work on the code and analysis together they 
  need to be able to share data to show findings to each other. This should
  be as automatic as possible to speed the development cycle.

\textbf{\tsix: improve cellPACK:}
  The developers constantly improve the quality and speed of
  cellPACK. In addition, the insight gained on the impact of particular parameters leads to removing some and adding others.

With our new approach, we aim to speed up their work, reduce communication
overhead and support a more systematic analysis of different cellPACK outputs.
Instead of single input - output scenarios, cellPACKexplore enables the setup
of an experiment (\tone) without programming knowledge using a visual interface
(see~\autoref{sec:inputScreen}) to sample a subset of parameters in a specified
range. This speeds up their workdflow as both developers can set up and run experiments 
now. Afterwards the ensemble of all outputs can be analyzed in a visual
interface (see~\autoref{sec:analyzeScreen}) in the old setup only one output cell could be analyzed after the other. 
cellPACKexplore also shows several
statistical metrics that are of interest to the developers. The interface of cellPACKexplore 
is adjustable to add new metrics in the future.
\autoref{fig:workflow} shows the workflow, while all the tasks had to be done
manually before, with cellPACKexplore we support \tone, \ttwo, \tthree~and
\tfour~visually. To reduce the communication overhead, we embed all the tasks
in one system accessible using a common web browser. This also supports \tfive, as the data is easily accessible. 
In the new cellPACKexplore interface, the cellPACK developers first configure 
an experiment (\tone) by selecting a cellPACK recipe, and input parameters
to analyze. cellPACK then generates hundreds of outputs and cellPACKexplore computes several derived metrics based on the outputs generated by cellPACK.
When all the computation is done, the cellPACK developer can explore the generated ensemble
(\ttwo, \tthree, \tfour) using cellPACKexplore's interface. After an analysis one can either setup a new experiment (\tone), 
discuss and share the results (\tfive) or modify
the cellPACK code (\tsix). Updating the cellPACK code is the only task not integrated in
cellPACKexplore and still done manually, since -- at this point -- this still requires a programmers expertise.

In the following, we outline some details we discovered analyzing the developer's model building task.

The main tasks we needed to support the
developers in was model validation% (\tone\ and \ttwo) 
(i.e. does cellPACK produce
correct cells compared to current domain knowledge). This is different to model usage, where 
the underlying system is usually validated and well tested.
A core task we found specific for the developers of
cellPACK is the need to identify crucial parameters that need to be exposed to
potential users of cellPACK vs. parameters that should not be exposed to the cellPACK user but
require reasonable default values. In a typical machine learning or other modelling scenario the focus lies on finding optimal parameters. Here, however, the focus is on parameters that show a great expressiveness as well as the biological connotation and are therefore useful for an interactive manipulation.

%
%\begin{figure}[htb]
% \centering
% \includegraphics[width=.99\linewidth]{cellpackSections/figures/timeline.png}

% \caption{\label{fig:timeline}
% In each of the two iteration cycles we started by a discussion about what to focus on in the interface and developed paper prototypes. During the implementation phases we got feedback from the cellPACK developers who tested the final tool.}
%\end{figure}
%-------------------------------------------------------------------------

\section{Methodology}
\label{sec:approach}

We followed the guidelines of the \textit{Design Study
Methodology}~\cite{sedlmair:2012} and iteratively refined cellPACKexplore over
the course of ten months. We first analyzed  the cellPACK developers, their data and tasks, and then
defined our design goals based on this analysis. In each of two main cycles we
started with several paper prototypes and presented them to the developers of
cellPACK. After getting feedback on the initial prototypes and refining them,
we implemented a high-fidelity prototype, enhanced and tested by the cellPACK developers on
their local work stations. During the whole project we have been in close
collaboration with the cellPACK developers.

During the first of these two cycles the cellPACK developers performed a limited number of experiments---analyzing one of their core test recipes that packs circles with different radii onto a 2D plane. Several meetings revealed their work to be more complex than initially observed. As they presented the capabilities of their code to us, it became clear that they were also interested in features of their output beyond the simple distribution of ingredients. For example, they incorporated a parameter that controlled the attraction of one sphere ingredient to another. Our initial prototype, which used an automatic clustering algorithm to group the outputs based solely on the distribution of ingredients, did not work well. The set of features required for clustering was not clear a priori and was changing from case to case. The cellPACK developers needed a tool that could provide meaningful clusters to clarify parameter effects like incorporating the binding probability between two ingredients. We found that the criteria for clustering changed frequently depending on the part of the packing algorithm tested as well as on the cell currently being developed. We consider these ongoing changes in the underlying data as a characteristic of \textit{model building}.
%that the developers analyzed their results on a variety of features that are prone to change with each new parameter added. 
Hence, we decided not to employ any particular clustering algorithm but instead to let the cellPACK developers interactively adapt groups of outputs by filtering the whole ensemble on several features (input parameters as well as output metrics) at once. These features are interchangeable easily if the data changes without requiring a new analysis process or a new interface design.

In the second iteration, we designed a new interface. This time we developed several statistical metrics to address the diversity of important characteristics in the output and also allowed the cellPACK developers to directly explore the 3D outputs via orthogonal projections. Considering the feedback and new challenges arising from the expansion to 3D, we again started developing and improving paper prototypes before implementing a high fidelity prototype. Again all design decisions have been made in close collaboration with the developers of cellPACK. The resulting tool has been tested by them (and iteratively refined) with some smaller experiments of approximately hundreds of runs. The initial evaluation phase revealed some minor issues which have been resolved before a second evaluation phase. To make cellPACKexplore easier to access for the cellPACK developers, we decided to install cellPACKexplore on a web-server and let the developers interact through a web-browser. They used the online version of cellPACKexplore to more exhaustively analyze their cells and to test the functionality of specific input parameters (see \autoref{sec:evaluation} for details).

%-------------------------------------------------------------------------
\section{Design}
\label{sec:design}

%-----------------------------------------
% input screen
\begin{figure}[tb]
  \centering
  \includegraphics[width=.99\linewidth]{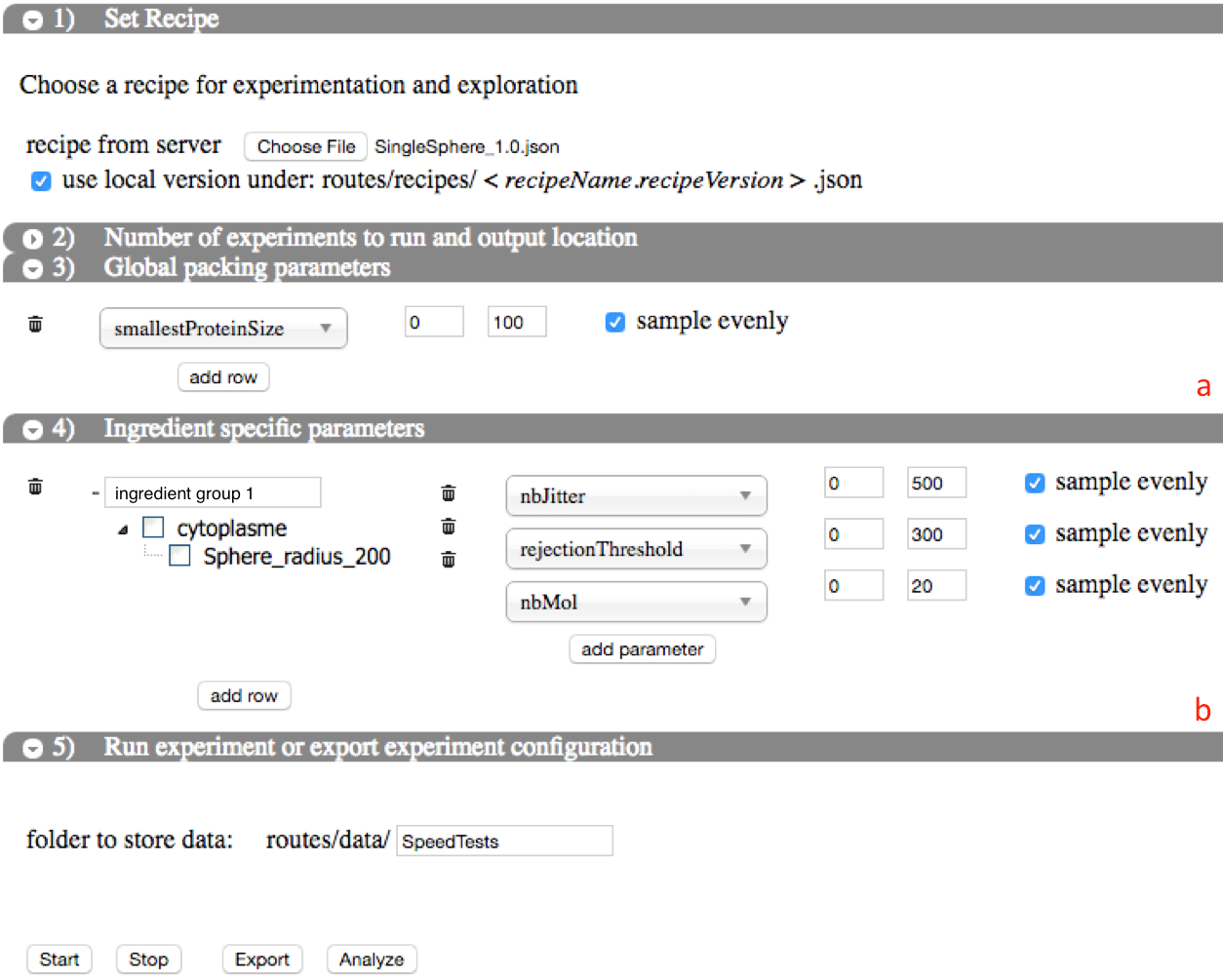}

  \caption{
  The input interface. Each vertical panel is one step in the setup of an experiment. From top to bottom: 1) recipe specification 2) cellPACKexplore settings (collapsed) 3) cellPACK general packing parameters 4) cellPACK ingredient parameters 5) start/export configuration. 
  % {\color{red} TM: open number 2 as well or is there a good reason not to? - yes. It would make the image way bigger and the section is just meta data - path locations etc.}
  }
    \label{fig:inputWhole}
\end{figure}
%-----------------------------------------

%-----------------------------------------
% analyze screen with annotations
\begin{figure*}[tb]
  \centering
\includegraphics[width=.99\linewidth]{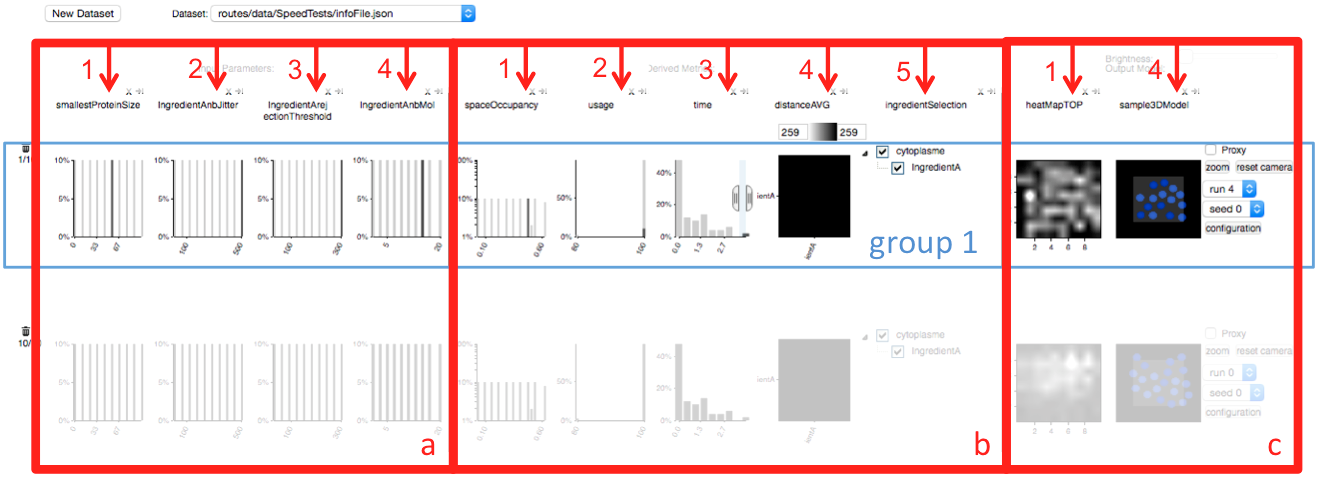}
  \caption{
           The analysis interface. Three main columns for a) input parameters b) derived statistical metrics c) spatial output presentation.}
    \label{fig:analyzeWhole}
\end{figure*}
%-----------------------------------------

cellPACKexplore is built as a client-server design with a web front-end. We chose this design as it simplifies installation and helped the developers of cellPACK to share (\tfive) their results and discuss findings. Plots are implemented using D3~\cite{bostock:2011} and 
crossfilter~\cite{crossfilter}
supports interactive filtering of large datasets. We used the approach of
Talbot et al.~\cite{talbot:2010} for axis labeling.

We separated the five tasks into two sequential interfaces (each making use of the full screen), one for the setup of an
experiment (\tone) and one for the analysis of the experiment results (task 2
through 5). This decision is based on the observation that the experiment
setup was done carefully and once an experiment is set up, most time was spent
on the analysis of the experiment, rarely reversing back to change the setup.
For a better understanding of our tool, in the supplemental material we provide a video demonstrating a walkthrough through cellPACKexplorer.

% input interface
%-------------------------------------------------------------------------
\subsection{Input screen}
\label{sec:inputScreen}

We developed the input screen~(\autoref{fig:inputWhole}) to support task \tone~of the developers. The cellPACK developers use this screen to select a cellPACK recipe file that specifies default values and ingredients to be packed. Then they specify parameters they want to explore in the experiment and set ranges they want to analyze.
There are parameters for each ingredient as well as parameters controlling the packing simulation, making the parameter space very large. As each ingredient has its own parameter configuration, the complexity increases with each added ingredient.
Due to this complex parameter space that changes depending on the selected recipe (and thereby the selected ingredients), the design of the input interface was challenging.

To address the complexity of an experiment setup and make it accessible for both developers (the traditional workflow was one developer providing custom python scripts on request of the other developer) we organized the setup of the experiment into five tasks, also reflected in the interface shown in \autoref{fig:inputWhole}.

{\bf 1) Set recipe:} At first the developer selects a cellPACK recipe he or she wants to analyze. This recipe is a \textit{.json} text file that specifies the packing volume, which ingredient types to pack, and default values for cellPACK input parameters. 
	
{\bf 2) Number of runs and output location:} The second step configures the experiment and determines how many cells should be computed ($N$ and $R$, see \autoref{sec:abstraction:data}).
	
{\bf 3) Global packing parameters:} In step 3, one sets general parameters of the simulation (see~\autoref{sec:cellpack}). The developer is only required to set the parameters they want to vary, other parameters remain at default values as specified by the selected recipe file. For each sampled parameter the cellPACK developer can also choose a sampling method. He can either select (deterministic) even sampling or stochastically uniform sampling. All parameters that are evenly sampled are formed through a multidimensional Cartesian lattice. This is ok for a small number of parameters but can lead to a combinatorial explosion quite quickly. Hence, it should be used with caution. 
%If he selects to sample evenly each parameter is sampled equally spaced in the range and all possible combinations of parameter values are computed. This space can get very huge and should only be applied for a small number of samples. If this option is not selected random values in the specified range are chosen to sample the parameter space.
	
{\bf 4) Ingredient specific parameters:} \textit{Ingredient parameters} require a more complex setup as each ingredient type can have its own set of parameter values (compare ingredient parameters~\autoref{sec:cellpack}).
	 When using the cellPACK itself, setting parameters is a very tedious process since each parameter value must be set by hand in a configuration file (recipe).
	 	To overcome this, we provide a searchable list of all available parameters.  When the cellPACK developer starts typing only parameters whose name matches are shown in the list. As we focused on the developers of cellPACK they know the names of parameters they want to sample and therefore can save a lot of time using this feature.
	 Because we observed the cellPACK developers changing the same parameter across multiple ingredients, we added the ability to modify parameter values for groups of ingredients. The tree representation of ingredients mimics the structure of ingredients in a cellPACK recipe. They can select ingredients in the tree and then select parameters and ranges to be sampled for the selected ingredients.	 
	 Once parameters for a particular group of ingredients are set, the cellPACK developer can create a new row (with a new tree) and specify parameters for a different subset of ingredients.
	
{\bf 5) Execute or export experiment:} Once the cellPACK developer has finished setting up an experiment, he can either run it directly on the server or download the configuration. The download option is helpful if he wants to run the experiment on another machine (i.e. with more computational power), send the setup to someone else (i.e. developers working together), or simply do the computation later.
After the setup of the experiment, the computation of the outputs is done offline and does not require any interaction so the cellPACK developers can concentrate on other work while the data is computed.
% (see \autoref{fig:workflow}, \tone/\emph{generate data}).

% analyze interface
%-------------------------------------------------------------------------
\subsection{Analysis screen}
\label{sec:analyzeScreen}

Once all the outputs are computed and derived statistical
metrics are ready, the analysis screen (\autoref{fig:analyzeWhole}) can be used to analyze the generated ensemble of cells.

Each row in the interface represents a subset of all runs of the selected experiment. This allows the cellPACK developer to compare (\ttwo) different subsets of the output ensemble. The plots for each row
can be divided into three logical groups: distribution of input
parameters~(\autoref{fig:analyzeWhole}a), distribution of derived
metrics~(\autoref{fig:analyzeWhole}b), and renderings of the
outputs~(\autoref{fig:analyzeWhole}c). The total number of runs summarized is
shown on the left side of the interface, where the developer can also add and
delete rows (select a new subset of the outputs). A new row initially starts with all runs of the experiment (i.e. no filters are applied and the whole output set is visible). The cellPACK developer can interactively adapt which runs are part of a horizontal group by
creating a filter on any one or several of the input or derived measure
histograms. The filters are combined with an AND operation such that groups are formed where each single output of a run has to fulfill all the filter constraints to be part of a horizontal row.  Each filter only influences its own horizontal row in the interface.
While manually filtering the runs can be cumbersome, it provides the greatest flexibility for exploring the output based on often changing objectives.
The column layout provides the flexibility to add other features (metrics) in the future or delete existing once without changing the interface. This is something we found out to be important for \textit{model building}.
We explain the different columns in turn.

\subsubsection{Input parameters}

The left-most columns (\autoref{fig:analyzeWhole}a) show one histogram for each
sampled input parameter. It supports the developer in understanding an input parameter's influence on the generated outputs (\tthree) and identify good default values for parameters (\tfour). Only sampled input parameters which have been selected by the developers in the input interface (\autoref{fig:inputWhole}) are shown. All others remain at default values for the whole ensemble.
% set and are not shown therefore. 
The horizontal axes of each graph represent
the sampled parameter value (numerical or categorical). The vertical axes
indicates the frequency, i.e. how often a specific value has been used to generate
outputs in that row. If the developer sets a constraint (horizontal selection
on any graph), all of the charts are updated accordingly. The full histogram for all runs remains transparent in the background as context. The updated histogram shows the distribution of the currently selected subset of a row.

\subsubsection{Derived metrics}
\label{sec:design:derived}

The center set of columns~(\autoref{fig:analyzeWhole}b) show histograms of various derived outputs. We created these columns to help the cellPACK developer to quickly identify subsets of interest without scrolling through output images one by one. E.g. we watched the developers looking for outputs where some ingredients failed to pack or outputs that took a very long computation time. The y axis on all histograms shows the frequency of each value on the x axis in the whole or currently selected dataset. This is consistent with the input parameter column.

As some graphs are computed on an ingredient basis, the cellPACK developer can focus on a
subset of ingredients by individually selecting or deselecting them in the
tree~(\autoref{fig:analyzeWhole}b5) also used on the input screen. This supports the analysis of a single ingredient, for example checking how much of the available space it covers.

Finding good derived metrics is difficult and hence, they are constantly revised. During the development of cellPACKexplore and as our understanding of the packing algorithm improved, we suggested a number of new metrics. The developers of cellPACK also requested a number
of different metrics and explored different parameters requiring new metrics.
As we progressed through the development of cellPACKexplore, we refined the list of output metrics. 
The column layout in the interface gave us the possibility to easily swap in and out metrics and add new ones.
In the current version of cellPACKexplore we show the following metrics:
\textit{spaceOccupancy}, \textit{usage}, and \textit{distanceAVG}. These metrics show derived geometrical properties of the cellPACK outputs:

{\bf spaceOccupancy:} The developers are interested in the concentration of different ingredients as this is crucial to assure biologically valid outputs. The \textit{SpaceOccupancy}~(\autoref{fig:analyzeWhole}b1) histogram shows the distribution of the percentage (horizontal axis) of the total packing volume covered by an ingredient type.
For example, if ingredient A takes up 50\% of the whole cube in which we are packing
then the \textit{SpaceOccupancy} value is 50\%. Within each run the occupancy
for each ingredient is averaged over all cells computed with different seeds. This
measure gives an idea of the concentration of ingredients compared to the total
volume, the denser the volume is packed, the higher this measure will be.

 {\bf usage:}
An important aspect to assure that cellPACK produces correct outputs is to have the full usage of ingredients, i.e. the number of copies of an ingredient type the cellPACK developer wants to pack should be equal to the number of copies that is actually packed in the generated cell.
In some cases these two numbers might differ. E.g. the ingredient might not find a place as there is not enough space left or its parameters 
do not allow certain positions. This would result in a usage below 100\%. The developers want
to identify these cells, and investigate them further. Ideally this histogram would only show one peak at a usage of 100\% (as is the case in the example of \autoref{fig:analyzeWhole}b2).

 {\bf distanceAVG:}
Within a cell proteins rarely act alone. Molecular processes are carried out by
    the interactions occurring between specific proteins. Moreover, the
    interior of cells is a crowded environment. This crowding effect can make
    molecules in cells behave in radically different ways than in test-tube
    assays. It is thus important to have a metric that can represent the
    crowding property of a given cell generated by cellPACK. To analyze
    this in cellPACKexplore, we developed the
    \textit{distanceAVG}~(\autoref{fig:analyzeWhole}b4) measure. It measures
    the distribution of pairwise distances between each ingredient instance to
    every other ingredient instance for each cellPACK output (also known as the
    radial distribution function in physics) averaged over the subset of a run.
    As each ingredient type is repeated multiple times in a typical packing we
    compute a distance matrix using the average distance between ingedients. 
    It is displayed as a heatmap (\autoref{fig:analyzeWhole}b4),
    mapping low distance to white and high distances to black. 
    In this figure, we have only one ingredient hence there is only a single distance in the matrix.

%  \item {\bf run-time:}
The  {\bf run-time}~(\autoref{fig:analyzeWhole}b3) of outputs generated in an experiment provides crucial information to enable
the developers to improve cellPACK's efficiency (\tsix). It shows which parameters have the
greatest impact on the computation time. Within an experiment it often happens 
that all outputs require approximately the same time except for one that takes much longer. Being able to filter on these outputs, the interface shows what input configuration caused the long computation time.
This metric can also be used to find a proper trade off between
a high density of packed ingredients and a reasonable computation time.
The developer can quickly assess the computation time and compare it to the achieved accuracy of the outputs. If the developer is interested in accuracy represented by how dense ingredients are packed in a cell he can use the \textit{spaceOccupancy} metric, showing how much of the space is occupied by an ingredient, to estimate the density. If he wants to check overlaps and intersections between ingredients he can make use of the \textit{distanceAVG} graph and compare pairwise distances of ingredient types.
Setting filters on these charts, cellPACKexplore can be used to quickly focus on a subset of outputs that satisfy special criteria. Subsequently, these outputs can be inspected in more detail on the right side of the row (\autoref{sec:design:packing}).

%-------------------------------------------------------
% figure explaining heatmap and distribution
\begin{figure}[htb]
  \centering
  \includegraphics[width=0.8\linewidth]{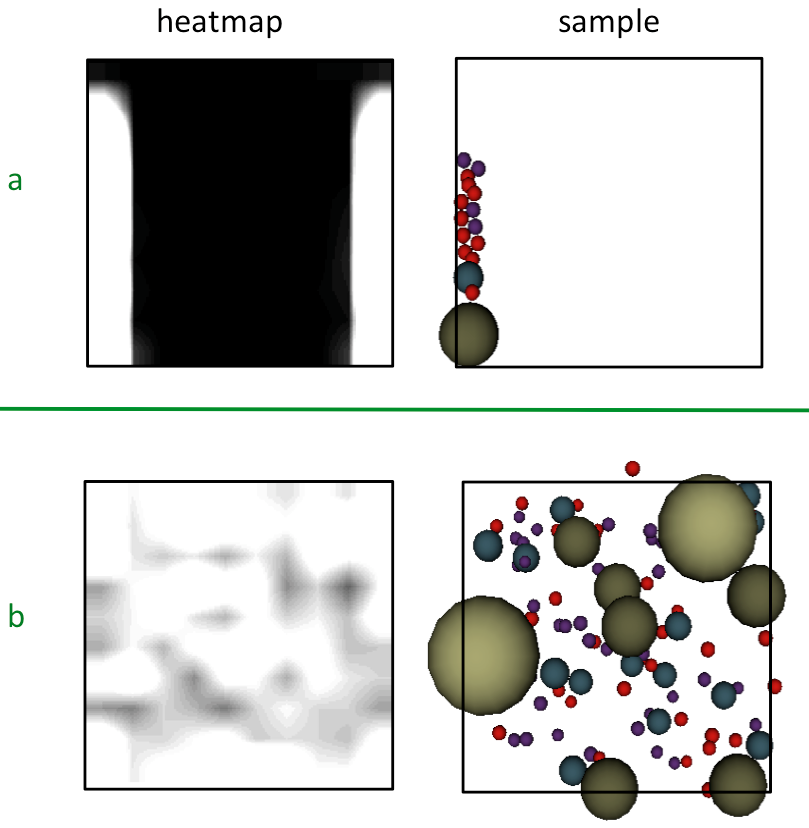}
  \caption{\label{fig:binning} Left: heatmap projected along the y-axis of the same ensemble. Filled space is colored white while empty space is black. Right: one sample of the ensemble set projected along the y-axis. Note the bias in the upper row (a) towards the left edge while the rest of the volume is empty (black). In the heatmap ingredients reaching out of the packing volume (black rectangle) periodically come back in on the opposite site (periodic boundary condition) which explains the white stripe on the right side in the upper row. The lower row (b) shows a random uniform distribution.
  }
\end{figure}
%-------------------------------------------------------

\subsubsection{Packing columns}
\label{sec:design:packing}

The right part of the interface~(\autoref{fig:analyzeWhole}c) gives the
cellPACK developers access to the direct output of cellPACK, which is a stochastic 3D volume.
The first image of~(\autoref{fig:analyzeWhole}c) shows the density of ingredients within the
probabilistic volume for different orthographic projections (top, right,
front). In~\autoref{fig:analyzeWhole}c, columns 2 and 3 have been closed by the developer as a 2D packing is analyzed. To compute these heatmaps the packing volume is discretized into a
user-defined number of subvolumes. For each of these subvolumes we
compute the volume covered by an ingredient type divided by the total volume of the
voxel. The resulting values are mapped to varying grey levels (black means an empty subvolume containing no ingredients). 
\autoref{fig:binning} shows
a comparison of a biased cellPACK output with a lot of empty (black) space (upper row)
and an output with uniform distribution (whole image is gray or white meaning that there are ingredients in all voxels) (lower row). We can see that there is a
bias towards the boundaries of the packing volume in the top row because the
border is brighter. Heatmaps have been used by the cellPACK developers before to analyze their cells. We chose to incorporate them in cellPACKexplore 
to provide access to their initially used analysis methods. In our experience these (direct) visual depictions of the cells are easier to understand and were preferred by the less technically trained of our two users (cellPACK developers).

The last column~(\autoref{fig:analyzeWhole}c4) shows an interactive 3D view of one cellPACK output of a run in that row and gives the developer the option to inspect details for specific cells.
The developer can interactively change which cell (run and seed) to present by selecting a different option in the dropdown menus. Outputs not part of the horizontal group are disabled. To interact with the 3D cell, the developer can use the mouse wheel to zoom and mouse dragging to rotate and translate the cell. We used billboard imposters for spheres to speed up the rendering. To further improve performance, in case of highly crowded cells (e.g. HIV), the developer can turn on the ``proxy'' option: each ingredient will be replaced by a single sphere encapsulating the original ingredient's shape. This representation shows the spherical proxy used by cellPACK to resolve the intersections between different ingredients while packing them to form the final cell.
To better analyze one specific cell, the whole view can be enlarged by a click on the ``zoom'' button. After testing the tool, the developers of cellPACK were interested in the exact parameter configuration of the cell presented in the viewer. Hence, selecting ``configuration'' opens a tooltip with detailed information about the sampled parameters that yield that specific output.

The interface can easily be adapted by closing or opening columns of information (e.g. when exploring a 2D recipe, only one of the projected heatmaps is needed, features can be shown/hidden depending on the analyzed parameter). After the exploration of an experiment, the underlying model (cellPACK) might be improved (\tsix) and a new experiment can be started.

%-------------------------------------------------------------------------
\section{Case Study}
\label{sec:evaluation}

cellPACKexplore was tested by the two developers of cellPACK, who were our target audience. They ran the tool directly in a 
web-browser and did not need to install any software. They used cellPACKexplore to validate the functionality of new parameters, optimize recipes and find a proper tradeoff between computation time and accuracy (achieved density of ingredients). They also ran some biological experiments asking ``What set of parameters of protein interaction can lead to observed patterns?'' In a qualitative assessment, the 
developers of cellPACK 
described cellPACKexplore as being extremely helpful in four major ways: 1) accessibility and 
workflow speed; 2) software (cellPACK) debugging; 3) recipe optimization; 4) hypothesis generation in 
biological research. After not looking at the code for several months, one developer used cellPACKexplore to remember what different parameters were doing and what values to choose. In the following we summarize their qualitative feedback in more detail.

First, being able to run cellPACK and to analyze cellPACK results on a web-server eliminated 
their constant hassle of maintaining and running cellPACK across the diverse collection of 
computers they use at home and at work (multiple operating systems, incompatible Python 
versions, etc). Being able to setup experiments and analyze the results from any machine, 
including a smartphone, was noted as a huge improvement in their workflow. 
Both cellPACK developers described significant speedups to their workflow. One 
developer describes how the experiments he typically designs take him 
between thirty seconds and three minutes for setting up and starting to use cellPACKexplore, whereas for the past several years, 
scripting and debugging the same experiment used to take him between twenty minutes and one 
hour. Similarly, the cellPACK developers described significant speedups 
for the analysis of cellPACK outputs, as many of the manual 
spreadsheet analysis tasks and the writing of custom analysis scripts to 
generate additional metrics on the cellPACK outputs are automated now. 

\begin{figure}[tb]
  \centering
  \includegraphics[width=.9\linewidth]{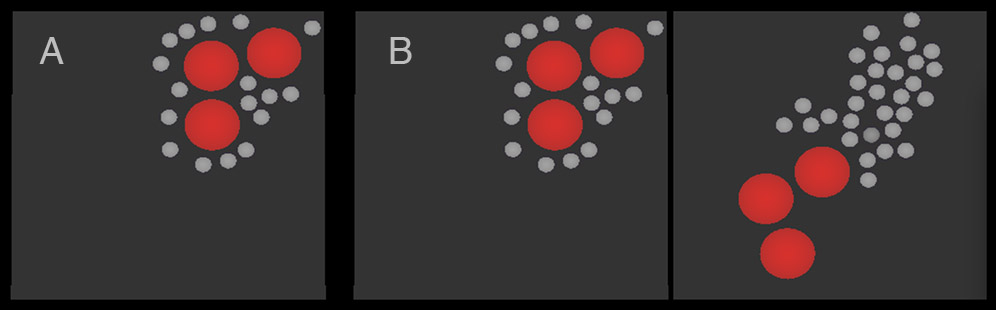}

  \caption{\label{fig:evaluationFig1}
  (A) The original recipe prior to adding the \emph{weight} parameter code shows how Ingredient2 (small gray spheres) always packed close to Ingredient1 (large red spheres.) (B) The new version of the original recipe shown in (A) has the \emph{weight} parameter added and cellPACKexplore has been used to sample the \emph{weight} probability from 0\% on the left (always bind to Ingredient1) to 100\% (always bind to Ingredient2) on the right.
 }
\end{figure}

Second, the interface 
enabled them to test new code for stability and functionality. For example, an 
ingredient parameter, \emph{weight}, that influences an ingredient's decision to pack 
close to a binding partner (another ingredient that has already been packed) was added by one of the developers. Without 
looking at the code, using only cellPACKexplore 
(on a cellphone), the other developer was able to quickly validate this new feature. He added 
the new \emph{weight} parameter to a known recipe, and sampled it in a range from 0-100\%  to confirm that at 0\% the results computed by cellPACK were the same as before 
(random as without the parameter, see \autoref{fig:evaluationFig1}A and \autoref{fig:evaluationFig1}B-left). 
As the probability was evenly dialed up to 100\% cellPACK 
produced results that matched the developer's hypothesis of how the weighting would influence IngredientB (see \autoref{fig:evaluationFig1}B-right).

%----------------------------
% figure 1
%----------------------------
One cellPACK developer also used cellPACKexplore for large-scale debugging tests and to isolate more subtle issues with 
newly added code. Instead of debugging heuristically by adjusting parameters 
and viewing one result at a time, cellPACKexplore enabled him to setup and analyze 
thousands of simulation runs at a time, which revealed statistical subtleties more 
readily. To give an example some packing outputs did not have a uniform distribution along the surface of a sphere as anticipated. Using cellPACKexplore the developers were able to see this bias and discovered that it happened due to non uniform meshing of the sphere surface using polar coordinates.

In a typical scenario the developer first runs small experiments and 
samples only two sets of parameters ($N=2$) at a time 
with typically just two seeds ($R=2$, \tone) to ensure the interface and the program are 
running correctly~(\ttwo). If the code failed, he discusses with his collaborator (\tfive) and/or he debugs it and 
adapts cellPACK~(\tsix). After resolving any preliminary issues he 
would run the whole experiment (sampling all parameters of interest) and test it with a small number 
of seeds ($R$, \tone) to confirm that the pairwise tested parameters all worked 
together (\ttwo). Finally, he greatly increases the number 
of seeds ($R$) for a deep analysis to explore 
the behavior of cellPACK (\ttwo, \tthree, \tfour). 
One of the developers usually starts looking at some of the sample outputs on the right side of the 
interface. Then he sets some filtering constraints on maxima (followed by minima) of the parameter's 
sampling range as he expects difficulties to appear at the extremes. After he narrowed down the search space 
he looks at the results in the right most column.
If the cellPACK output is 
incorrect he adapts cellPACK again (\tsix) and starts a second analysis iteration.

\begin{figure}[tb]
  \centering
  \includegraphics[width=.99\linewidth]{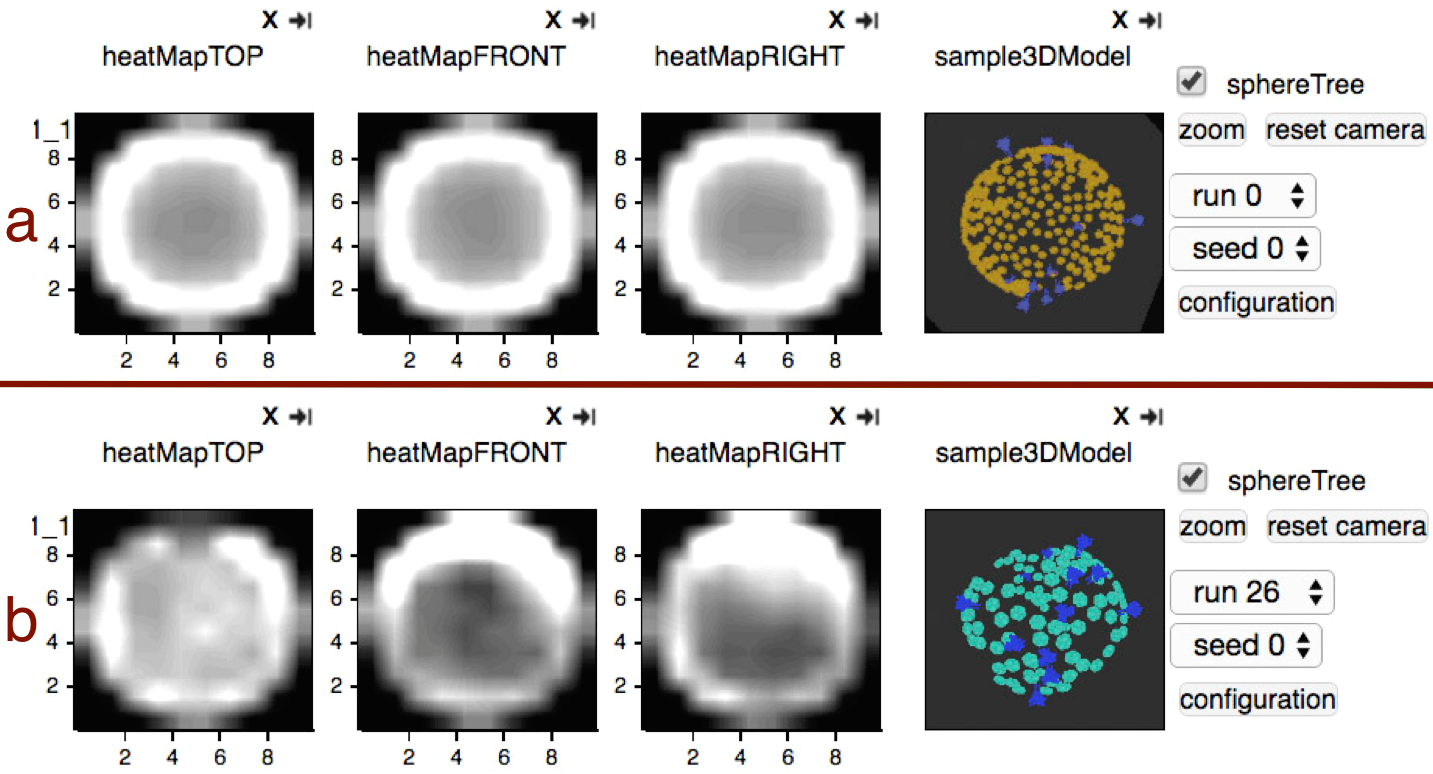}

  \caption{\label{fig:evaluationFig2}
  An example of how a subtle bug was found in the cellPACK 
  code that was written to pack objects close together on the surface of a sphere 
  100\% of the time. The heatmaps quickly revealed a uniform random 
  distribution on the spherical surface~(a). The core code was adapted and a second experiment revealed the anticipated hotspots at some locations on the surface~(b).
  %The heat maps quickly revealed that the recipe was not using the constraint to pack objects close to each other and was reverting to the random distribution default~(a).
}
\end{figure}

Using this exhaustive approach, and a wide sampling range, the developers of cellPACK 
observed some problems that could not have been noticed with the smaller experiments 
they were doing before. Manually scrutinizing hundreds of 3D volumes using their old 
approach was time consuming and prone to error. cellPACKexplore's 
filtering options helped them to quickly discover issues such as repetitive/identical 
outputs or incorrect distributions that resulted from errors in the core 
code. Another issue were input parameter configurations that caused ingredient constraints (e.g. 
using a gradient or specifying that two ingredient types should pack close to 
each other) to be ignored or applied incorrectly. \autoref{fig:evaluationFig2} demonstrates one example of a difficult bug to spot manually that was relatively easy to find with cellPACKexplore.

%----------------------------
% figure 2
%----------------------------

Both cellPACK developers who tested cellPACKexplore found the derived metrics 
in the analysis interface~\autoref{fig:analyzeWhole}b (\autoref{sec:design:derived}) to be tricky to understand and confusing at a first glance. 
We addressed these challenges with online meetings and added 
tooltips explaining the charts which made some of the graphs easier to understand.
However, some metrics remained difficult to understand, 
even after we explained them multiple times. In response, we removed confusing metrics and developed new ones that are easier to interpret.
The developer who has a background as a scientific illustrator rarely uses 
the metrics column at all. He relies more on visual 
features directly linked to the outputs than on abstracted graphs. Although the metric charts revealed information in some experiments (showed patterns indicating e.g. a low or high concentration in connection to specific input parameter values) the visually working cellPACK developer still preferred to inspect the 
outputs in the 3D viewer directly.

The developers noted that the 
filtering options were well 
designed and that it helped them to narrow down the effect of particular parameters 
as well as combinations of parameters and their ranges. They could clump relationships on the metrics 
and the analysis interface enabled them 
to integrate and summarize any number of cells in a relatively easy manner. 
One developer noted that he often tests cellPACK and cellPACK recipes for their ability to generate 
the default standard packing of a uniform random distribution.
He used the intensity of the heatmaps (even intensity across the heatmaps) as indicator 
of \emph{uniformity} to search parameter sets that generated uniform random distributions.

cellPACKexplore was considered useful for optimizing the performance 
of some test recipes in cellPACK. A common task for the 
scientific illustrator, is to pack lots of objects (e.g. molecules) densely together. 
Depending on the ingredient parameters,
% (compare~\autoref{sec:cellpack}, ingredient parameters), 
some cellPACK recipes have been found to pack ~95\% of 
the objects quickly and to then spend several hundred times longer to pack 
the remaining 5\%. When not being used for research, e.g. for generating illustrations to 
more general audiences, 95\% is often good enough to convey the message. In a simple 
test case, the developers of cellPACK attempted a dense packing and were 
quickly able to filter out the unreasonable (expensive) packing times and to see which 
input parameter configurations tended to cause the longer packing times.

%----------------------------
% figure 3
\begin{figure}[tb]
  \centering
  \includegraphics[width=.99\linewidth]{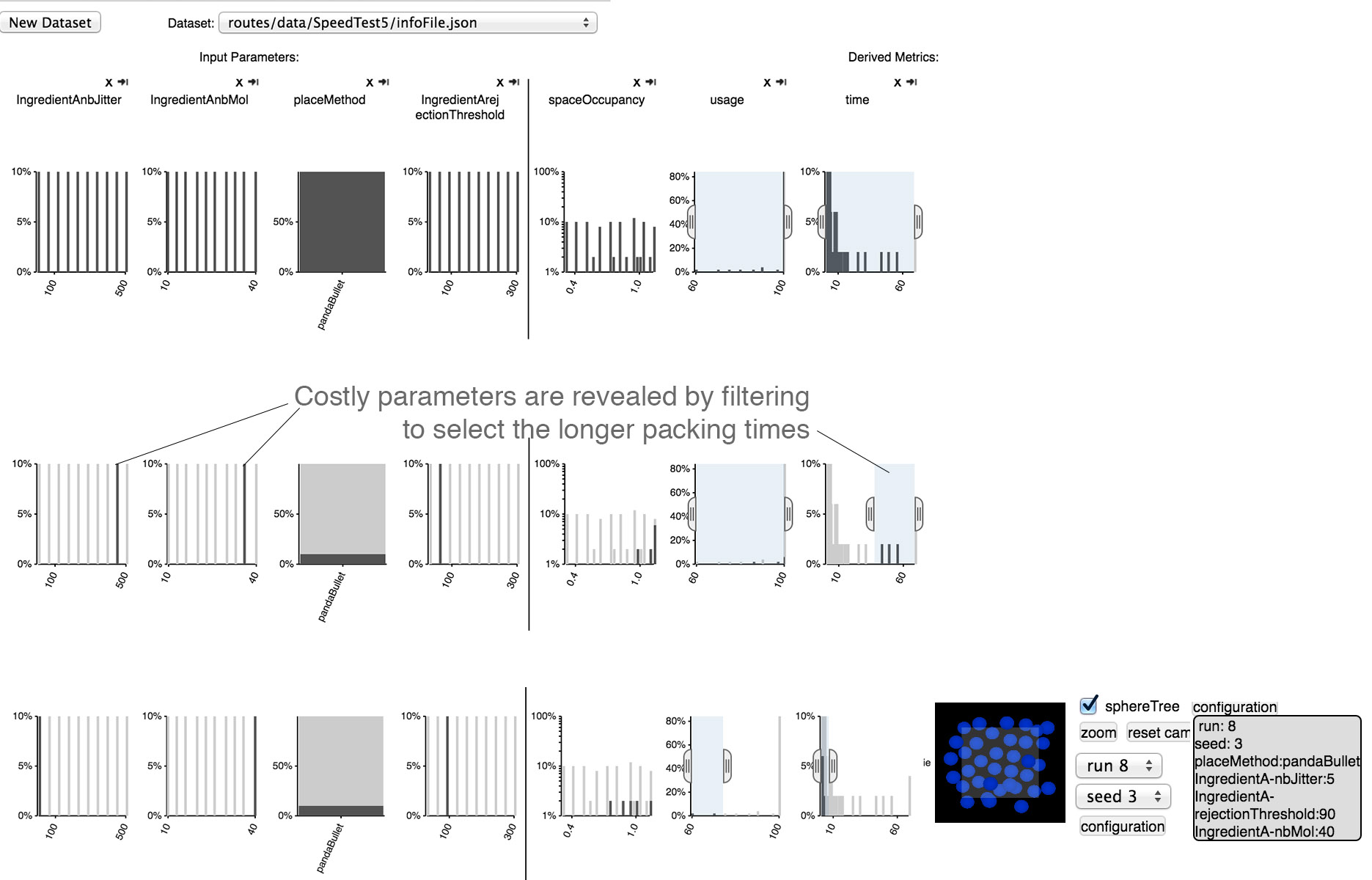}

  \caption{\label{fig:evaluationFig3}
  Top row shows the unfiltered results of a recipe that attempts to pack more spheres than possible into a plane (trying 10 to 40 spheres, approximately 30 can fit). Two parameters %that affect how many attempts a sphere makes before giving up
   were sampled: \texttt{nbJitter}: the number of attempts an ingredient makes to pack itself into a selected area before giving up if it is colliding (varied from 5 to 500); and \texttt{rejectionThreshold}: the number of failed jitters before giving up the entire attempt to pack any more copies of an ingredient (sampled from 30 to 300, N=10, R=5). The middle row shows filtering on time. This reveals that higher nbJitter causes a time increase. The bottom row shows how filtering on metrics helped to find a parameter setup packing the max number of spheres in the shortest time.
  %, e.g. nbMol=x, jitterMax=x, and rejectionThreshold=x.
}
\end{figure}
%----------------------------

%Both cellPACK developers requested us to add more options to control the sampling 
%strategy for input parameters. They often found it difficult to predict how many runs ($N$)
%would be required to sufficiently understand how the selected parameters would 
%interact. They found out that sampling the space with a full-factorial approach 
%was prohibitively expensive and often unnecessary. One developer requested a 
%per-parameter combinatorial option that could allow him to deeply sample some 
%parameters while skimming through others. We leave this for future work.

\begin{figure}[b]
  \centering
  \includegraphics[width=.99\linewidth]{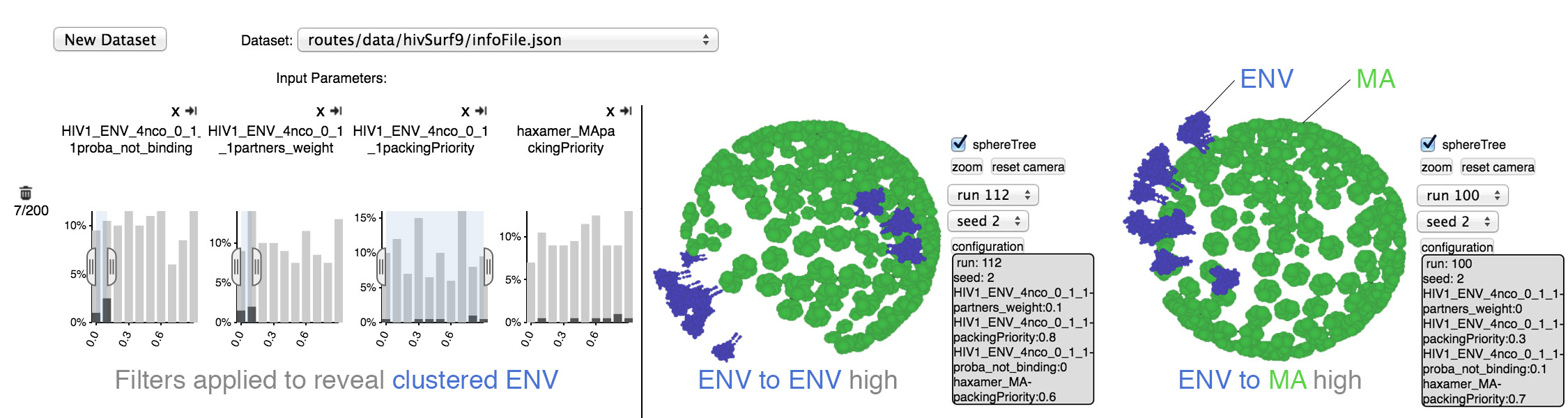}

  \caption{\label{fig:evaluationFig4}
  Strong ENV-ENV affinity parameters and strong ENV-MA parameters generate cells with polarized ENV clusters similar to the detailed models the cellPACK developers had generated previously with a more complex approach~\cite{johnson:2014}.}
\end{figure}

Lastly, the developers of cellPACK see great potential in using cellPACKexplore 
as a data exploration tool for hypothesis generation. They ran a few 
trial experiments to simulate Human Immunodeficiency Viruses (HIV) \cite{johnson:2014}
 with the new binding parameter \emph{weight} (in the previously described 
 scenario they tested the \emph{weight} parameter by packing spheres on a plane). In 
 cellPACKexplore the developers could visualize that the agent-based molecular 
 interaction parameters alone were able to generate cells, which by eye, and by the 
 general measures available in the interface, appear to mimic cellPACK outputs 
 that they had published using a more complex approach two years prior (see \autoref{fig:evaluationFig4}). They are 
 eager to thoroughly test these new approaches and have requested us to add their fluorescence 
 microscopy simulation tool~\cite{johnson:2015}
  to cellPACKexplore to enable a full evaluation. They plan to work with a biologist 
  collaborator to generate hypotheses for HIV surface protein molecular interactions that can be 
  tested in a wet lab.

In the end, however, the biggest achievement of cellPACKexplore was perhaps
the fact that the scientific illustrator person of the development team could
finally, and for the first time, run experiments without requiring his
technical colleague to setup, run, and prepare the results of an experiment.
This was very liberating and changed the dynamic within the development team.

%----------------------------
% figure 4 
%----------------------------

%-------------------------------------------------------------------------
\section{Discusssion}
\label{sec:discussion}

%\ttwnote{Here I think it would be good to talk more about other applications
%of cellPACKexplore. What you've written here is a really nice summary of
%the paper that I think would be a great conclusion section. But in this section
%talk about the wider applications of cellPACKexplore and relate the solutions
%of the system to other systems as well as the tasks you laid out in the 
%beginning} \tmnote{to Tom: it would be great to talk about application of cellPackExplore here -- what do you have in mind? However, I do find the more lengthy discussion of contributions and lessons learned quite valuable. It is too long for a conclusion section. A conclusion section should be short and to the point. no?}

The development of cellPACKexplore allowed us to work on refining the design
guidelines for building parameter space exploration systems. Specifically, we
can further explain the difference and similarities between systems for \emph{model building}
and \emph{model usage}. In some ways the requirements are similar because
iterative model building requires running the model. However, we identified
some additional requirements.
One of the most important insights is the focus on changing parameters. In \textit{model usage} parameter values are updated in \textit{model building} the parameters themselves are subject to change. Parameters might be added, deleted or modified. A difficult decision here is to expose enough parameters to future cellPACK users to let them build a variety of outputs but not overwhelm them with an unnecessary large parameter space. A parameter should also be interpretable by a future user of cellPACK (i.e. have a meaning in the domain) and have a specific influence on the output.

Our second major insight is the importance of objective measures -- the need to easily incorporate new ones, to help validate them and the ability to remove them if they have shown to be too complex or not useful. In our first prototypes we did not have the last column of the interface showing the 
packing result in a 3D viewer. The developers requested the option to see 
their data as they were used to before. We found out that this was very important 
for them to better understand the new metrics and work with the new interface. One 
of the developers rarely used the metrics at all and relied more on the direct 
representation of the output cells which is related to his background in 
scientific illustration.
In our meetings we realized that the developers of cellPACK had difficulties in 
understanding a metric we later removed from the interface. It showed the \emph{distribution} 
of ingredients in the volume in a barchart. In their previous analysis, they also used barcharts 
but with a different meaning to the x-axis. We learned that proposing a similar graph with changed axis did 
not work for the cellPACK developer who has not had a formal mathematical training.
%did not have a deep math background. 
We also realized 
that they focused more on metrics such as the \emph{run-time} which are easier to interpret for 
them. When using cellPACKexplore they usually used the filters on the input parameter 
histograms and inspected the results in the 3D viewer. They better understood 
some of the derived metrics when looking at their cell outputs next to them. 
Hence, in our experience, it will be crucial to find flexible ways to incorporate different, 
new, easy-to-interpret derived measures to help people (with or without mathematical background) work with large amounts of complicated model output (probabilistic ensembles in our case).

Further, dealing with ensembles of probabilistic volumes showed challenges. 
We tried to automatically cluster the output ensemble in the first iteration
but learned that this can be a hindrance if there is no clear metric to
cluster on. Omitting features to the clustering algorithm can hide important
aspects in the data. The possibility to partition the output ensemble
interactively enabled the cellPACK developers to adapt the grouping mechanism as they saw fit
% groups fitting the actual experiment and provides much more 
giving them the needed flexibility for their analysis. We imagine
that this is another aspect of the \textit{model building} tasks they are performing. Features of interest when analyzing the outputs are not clear at the beginning and might change during the development of the model (in our case cellPACK).

A further requirement we discovered is the need 
for an adaptable interface and the ability to quickly prototype.
cellPACK is still under active development. It is not possible to
predict what the developers will need to perform their analysis on in the future.
Our interface is adaptable as it is not tied to any specific
evaluation metric of the cellPACK cells. We believe that our
interface provides exactly this flexibility and the general approach we are 
using for the layout (columns for feature charts and rows for subsets of the data)
can also be used for other \textit{model-building} analysis tools.

An important detail we do want to mention here is that making cellPACKexplore
easily accessible (via web-browser) was very helpful for the final evaluation
of the tool, as the cellPACK developers did not have to worry about
installation issues. We could also quickly incorporate new metrics or code
changes and deploy them without worrying about installation issues.

%-------------------------------------------------------------------------
%\section{Limitations and Future Work}
%\label{sec:limitations}

%The design of cellPACKexplore was focused on supporting the developers of cellPACK. In the future we want to extend our impact to include the main future users of cellPACK: biology researchers and scientific illustrators. Although the current version uses parameter names that only experienced users (as the developers of cellPACK) can understand, a simple skin that adapts the parameter names to descriptive terms could quickly make cellPACKexplore accessible to other users as well. We believe that our current approach focusing on the developer's tasks is a suitable first prototype for a design of such a tool.

\section{Future Work}
\label{sec:Future Work}

% However we found out that there are still some improvements required to make cellPACKexplore even more useful in the future. 

While cellPACKexplore is a great support in the analysis of simple packings the scalability to realistic biological cells remains future work. While we achieved a scaling up from cells with 2-3 ingredients to tens of ingredients, a cellPACK output can consist of thousands of ingredients (e.g. E.Coli has about 1 million proteins made up of about $4,000$ unique proteins, Mycoplasma Mycoides is formed of about $50,000$ proteins made up of about $800$ unique proteins). This will require a change in our interface, as e.g. the tree representation for ingredients would not scale properly if all ingredients are unfolded.

The developers of cellPACK have requested the ability to select other derived metrics in the center column (\autoref{fig:analyzeWhole}b) ans the possibly to upload their own derived measures. We see a lot of potential future work in developing new metrics that reveal other features of the ensemble set and can be interpreted visually as easy as the outputs. We further realized that as the cellPACK outputs inspected changed (initially ingredients were packed in a box, in later experiments ingredients were packed on the surface of a sphere) some of our proposed metrics required an adaption. The \textit{spaceOccupancy} and the heatmaps on the right will not support analysis in case of a recipe that packs ingredients on the surface of a sphere.

A possible automatic clustering of similar probabilistic 3D cells could also help in scaling the interface to experiments with much larger runs, helping in scaling up to more parameters. While we deliberately decided not to use automatic clustering in this version of cellPACKexplore, we think of implementing a hybrid version in the future. The interface should initially display pre-computed clusters (horizontal rows) which the developers then can adapt later using the same filters as in the current version. We see a lot of potential work in finding a proper clustering method for the volumetric probabilistic dataset.

Finally we want to experiment with different sampling strategies. The cellPACK developers currently prefer the 'full combinatorial' method, that creates all possible combinations for the sampled input parameters (after discretization of the continuous parameters) this approach does not scale well if the sampling range or number of parameters sampled increases. We also anticipate to provide some guidance in choosing a proper sampling strategy and number of samples required to get a statistically meaningful output.

%-------------------------------------------------------------------------
\section{Conclusion}
\label{sec:conclusion}

To the best of our knowledge, this work, for the first time in the visualization literature, is presenting a task analysis of a \textit{model building} process in the biological domain. We specifically focus on the development of cellPACK for generating complex virtual cells that are difficult to parametrize and validate. We compared and contrasted the challenges and tasks performed by the cellPACK developers, related to \textit{model building} and \textit{model usage}. Specifically, we identified the need to add, remove, and find proper defaults for parameters guiding the modelling process as a novel task to be performed. Further, the ability to incorporate and validate new derived measures proved crucial and difficult for the success of the modelling pipeline.
Based on this breakdown we created cellPACKexplore supporting the developers in analyzing input parameter effects on outputs as well as the distribution of objects in a volume for a probabilistic volume ensemble dataset.

\iffalse
\acknowledgments{
The authors wish to thank A, B, C. This work was supported in part by
a grant from XYZ.}
\fi

%\bibliographystyle{abbrv}
%\bibliographystyle{abbrv-doi}
%\bibliographystyle{abbrv-doi-narrow}
%\bibliographystyle{abbrv-doi-hyperref}
\bibliographystyle{abbrv-doi-hyperref-narrow}
%%use following if all content of bibtex file should be shown
%\nocite{*}
\bibliography{merged}
\end{document}